\documentstyle[preprint,aps,prb,eqsecnum]{revtex}

\title{The FIR-absorption of short period quantum wires and the
       transition from one to two dimensions}
\author{Andrei Manolescu$^*$}
\address{Institutul de Fizica \c{s}i Tehnologia Materialelor,\\ 
         C.\ P.\ MG-7 Bucure\c{s}ti-M\u{a}gurele, Romania. }
\author{Vidar Gudmundsson}
\address{Science Institute, University of Iceland, Dunhaga 3,
         IS-107 Reykjavik, Iceland.}

\tighten

\input epsf

\draft

\begin{document}

\maketitle

\begin{abstract}
We investigate the FIR-absorption of short period parallel 
quantum wires in a perpendicular quantizing magnetic field. 
The external time-dependent electric field is linearly polarized
along the wire modulation. The mutual Coulomb interaction
of the electrons is treated self-consistently in the ground
state and in the absorption calculation within the Hartree
approximation. We consider the effects of a metal gate grating coupler, 
with the same or with a different period as the wire modulation,
on the absorption. The evolution of the magnetoplasmon
in the nonlocal region where it is split into several
Bernstein modes is discussed in the transition from:
narrow to broad wires, and isolated to overlapping wires.
We show that in the case of narrow and not strongly modulated
wires the absorption can be directly correlated with the
underlying electronic bandstructure.  
\end{abstract}

\vspace {7cm}

$^*$ Regular Associate of the International Centre for Theoretical Physics,
Trieste, Italy.


\newpage

\section{Introduction}
Parallel quantum wires can be produced in Al$_x$Ga$_{1-x}$As-GaAs
heterostructures by a periodically modulated metal gate 
on top of the Al$_x$Ga$_{1-x}$As insulator. By applying a voltage bias 
between the gate and the two-dimensional electron gas (2DEG) at the
insulator-semiconductor interface the properties of a 2DEG
can be studied in the entire range from a homogeneous system to isolated
wires. 

The theory of far-infrared (FIR) absorption of the {\em homogeneous}
2DEG in the presence of the metal gate acting as a {\em grating coupler}
has been developed by Zheng,
Schaich, and MacDonald\cite{Zheng90:8493} and by Liu and Das Sarma
\cite{Liu91:9122}.  In those papers the FIR response has been calculated 
semiclassically, allowing the presence of a weak magnetic field.
The FIR absorption of a {\em modulated} system, for instance due to the 
presence of a gate potential, has been extensively studied experimentally 
for a simple unidirectional modulation\cite{Hansen92:257} and calculated 
self-consistently for a sinusoidal potential and strong magnetic fields.
\cite{Wulf90:3113,Wulf90:7637} Modes specific to more structured modulation 
like parallel pairs of wires have been identified as well.\cite{Frank97:1950R}

The well known magnetoplasmon dispersion law for a homogeneous 2DEG
in a strong magnetic field is
\begin{equation}
      \omega^2=\omega_c^2+2\pi n_s e^2 q/(\kappa m)\,,
\label{mdlq}
\end{equation}
which is valid in the lowest order in the wave vector $q$, i.\ e.\ 
as long as the second term is much smaller than the first one.
For shorter wavelengths or for lower magnetic fields the magnetoplasma
oscillations have higher branches, around harmonics of the cyclotron
frequency, $n\omega_c, n=2,3,...,$ known as Bernstein modes.
\cite{Bernstein58:10} The Bernstein modes are likewise known in
single wires and dots.\cite{Gudmundsson17744:95,Brataas96:4797}

In this publication we want to focus our attention on the FIR absorption of
a periodically modulated 2DEG in a perpendicular quantizing magnetic field, 
in the regime where the Bernstein modes are a prominent feature of the 
frequency dispersion. Furthermore, we take into account the grating coupler 
effects of the metal 
gate on the incident FIR radiation.\cite{Wulf94:17670,Batke85:2367}
The external time-dependent electric field is hence modulated with  
a large wavevector, the lowest inverse lattice vector of the gate
structure. This is an approximation to the leading order compared to
the studies of the grating coupler effects mentioned 
above.\cite{Zheng90:8493,Liu91:9122}
In principle, the 2DEG can be modulated by an independent method,
i.\ e.\ not by the gate structure. 
 
We calculate the FIR absorption fully quantum mechanically with the
help of the inverse dielectric function. The mutual Coulomb interaction 
of the electrons  is treated within the Hartree approximation both in the 
ground state and in the absorption calculation, i.\ e.\ the absorption is 
obtained in the framework of the random phase approximation (RPA). The 
intra- or inter-wire Coulomb interactions thus enter the model on equal 
footing.

We show that as the width of the wires is reduced the resulting 
Landau bandstructure can be identified through the effects of its
van Hove singularities (vHS) on the FIR absorption.
The evolution of the FIR absorption is traced from the case of
weakly coupled wires to the case of a weakly modulated 2DEG
by increasing the electron density but keeping the modulation
strength constant. 
We systematically use the induced density to sort out the
complex hierarchy of absorption modes caused by the modulation
of the electron gas or by the electric coupling between wires.
\section{Model}
We consider a modulated 2DEG located in the plane
${\bf r}=(x,y)$, and describe it with the Hamiltonian
\begin{equation}
      H=H_0+V_{\rm mod}+V_H\,.
\label{hamil}
\end{equation}
$H_0=({\bf P}+e{\bf A})^2/2m$ is the Hamiltonian of the
noninteracting system.  We employ the Landau gauge for
the vector potential, ${\bf A}=(0,Bx)$. $V_{\rm mod}(x)=V\cos (Kx)$ is
the simplest model of an electrostatic potential modulation
varying in only one direction. $V_H$ is the Hartree potential
felt by each electron, self-consistently with the particle
density,
\begin{eqnarray}
      &&\rho(x)=\sum_{p\ge 0} \rho_p \cos (pKx) \,,
\label{densf}\\
      &&V_H(x)=\frac{e^2}{\kappa}\int d{\bf r'}\frac{\rho(x')}
      {|{\bf r}-{\bf r'}|}\nonumber\\
      &&=\frac{e^2}{\kappa}\frac{2\pi}{K}
      \sum_{p\ge 1}\frac{\rho_p}{p} \cos (pKx) \,,
\label{harpo}
\end{eqnarray}
where $\kappa$ is the dielectric constant of the semiconductor host of the 
2DEG. We assume the electrical neutrality to be 
ensured by a uniform background 
of positive charges, of density $-\rho_0$.  The eigenstates of 
the Hamiltonian (\ref{hamil}) are calculated 
numerically, within an iterative scheme, by 
successive diagonalization in the basis of the wave functions corresponding 
to $H_0$, known as Landau wave functions.  The resulting single particle 
energies, $E_{n,X_0}=E_{n,X_0+a},\, a=2\pi/K$, have a one-dimensional
periodic band structure, where $n=0,1,2,...$ is the Landau quantum number 
and $X_0$ is the center coordinate.  We assume spin degeneracy. The 
eigenfunctions of $H$ have the form 
$L^{-1/2}\exp (-iX_0y/l^2)\psi_{n,X_0}(x)$, 
where $L$ is a normalization length and $l=(\hbar/eB)^{1/2}$ is the magnetic 
length.

We consider a time-dependent electric field incident to the 2DEG, 
linearly polarized in the direction of the modulation, $x$, with only
one Fourier component, of wave vector ${\bf q}=(q,0)$ and frequency $\omega$.  
This field is supposed to be sufficiently small, within the linear response 
regime.  We denote by $\phi_{\rm ext}(q,\omega)$ the associated electric 
potential.  

In order to evaluate the absorption due to plasma oscillations we need 
to calculate first the dielectric matrix $\varepsilon_{GG'}(q,\omega)$,
where $G$ and $G'$ are vectors in the reciprocal space, of the form
$mK,\, m=0,\pm 1,\pm 2,...$\,.  We use the random-phase approximation, 
as described e.g.\ in the paper by Wulf et.\ al., \cite{Wulf90:3113} in which 
$\varepsilon_{GG'}(q,\omega)=\delta_{GG'}-
\frac{2\pi e^2}{\kappa | q+G |}\chi_{GG'}(q,\omega)$.  The dielectric 
susceptibility $\chi_{GG'}$ is given by the Lindhard formula, which in our 
case reads
\begin{eqnarray}
      &&\chi_{GG'}(q,\omega)=\frac{1}{a\pi l^2}
      \sum_{nn'}\int_0^a dX_0\nonumber\\
      &&\times\frac{{\cal F}(E_{n,X_0})-{\cal F}(E_{n',X_0})}
      {E_{n,X_0}-E_{n',X_0}-\hbar\omega-i\eta}\nonumber\\
      &&\times\, {\cal J}_{nn';X_0}(q+G) {\cal J}^*_{nn';X_0}(q+G')\,,
\label{linfo}
\end{eqnarray}
where ${\cal F}$ is the Fermi function, $\eta\rightarrow 0^+$, and
${\cal J}_{nn';X_0}(q)=\langle \psi_{n',X_0}|e^{iqx}|
\psi_{n,X_0}\rangle$.

The absorbtion power can be calculated from the Joule law
of heating, which may be written as \cite{Dahl90:5763}
\begin{eqnarray}
      &&P(q,\omega)=-\frac{\omega}{4\pi}  
      {\rm Im} \, \varepsilon^{-1}_{GG} (q_1;\omega)\nonumber\\
      && \times \, q| \phi_{\rm ext}(q,\omega)|^2 \,,
\label{abspo}
\end{eqnarray}
where $q=q_1+G$. 
Due to the periodicity of the system $\varepsilon_{GG'}(q,\omega)=
\varepsilon_{G-K,G'-K}(q+K,\omega)$, and we can take $0\le q_1\le K$. 
We will be mostly interested in the case when the wavelength of the 
incident field is identical to that of the modulation, as in the absorption 
experiments with grating-coupler devices, that is $q=q_1=K$ and $G=0$.  For 
simplicity, we will  normalize the external potential in Eq. (\ref{abspo}) 
such that $\frac{1}{4\pi}q|\phi_{\rm ext}(q,\omega)|^2=1$.

In order to avoid calculation of the plasma poles of the dielectric 
matrix, we assume a finite $\eta=\hbar\omega_c/50$,
in Eq. (\ref{linfo}).  In other words we assume an unspecified dissipation
mechanism, which leads us directly to the oscillator strength of the 
collective modes, and thus to measurable results.

\section{Results}
The parameters of our model are those for GaAs, the electron mass 
$m=0.067m_0$ and $\kappa=12.4$.  The temperature will be fixed to 1 K.
We truncate the Fourier series such that $\mid mK\mid \le MK$,
and we achieve numerical convergence for $5 \le M \le 15$.

As has been mentioned in the introduction the magnetoplasmon dispersion
at low magnetic fields and short wavelengths splits into Bernstein
modes due to the interaction with harmonics of the cyclotron resonance

In Fig. 1a we show the absorption power in the regime of the Bernstein modes, 
for the homogeneous system, for wave vectors varying from $0.01K$
to $K$, with $K=2\pi/a$, and $a=50$ nm.  The electron density is fixed to 
$\rho_0=1.8\times 10^{11}$ cm$^{-2}$ and the magnetic field is $B=3.74$ T, 
which corresponds to the filling factor $\nu=2$. 
Eq. (\ref{mdlq}) holds only for wave vectors below $q=0.2K$
for the present choice of parameters.  
In Fig. 1b we display the results for a modulated 
system, with a modulation period $a=50$ nm, 
and a modulation amplitude $V=5$ meV.
As we observe, each peak corresponding to a Bernstein mode 
splits into several subpeaks.  The same modulation effect is evidenced 
in Fig. 1c, where we have fixed the wave vector of the incident field,
$q=K$, but we sweep the magnetic field such that the filling factor
varies between 1 and 3.  The traces for $\nu=1.2$ and $\nu=2$ are magnified 
in Fig. 2.  Each Bernstein mode has up to four subpeaks.

We start to analyze this complicated internal structure, by showing
in Fig. 3 the energy spectra corresponding to Fig. 2.
The periodic Landau bands are displayed in half of a Brillouin
zone, i.e. for $0\le KX_0\le \pi$.  Due to the short period, in the 
absence of the Coulomb interaction the energy bands are not parallel. 
This is seen here only for $\nu=2$, where the Fermi level is in an
energy gap, and thus the screening effects are small and the energy 
dispersion large, Fig. 3a. For $\nu=1.2$, Fig. 3b, the Hartree
screening reduces the energy dispersion to very narrow bands.

The absorption peaks of Fig. 2 reflect the nonparallel Landau bands,
or, equivalently, a certain dispersion, in the center-coordinate space,
of the energy interval between pairs of Landau bands.  This is 
best seen for $\nu=2$, where for each Bernstein mode we can energetically 
relate the lowest and the highest subpeaks to the inter-Landau-level 
transitions around the center and around the edges of the Brillouin zone,
i.\ e.\ $X_0=0$ and $X_0=\pi/K$ respectively. The energy dispersion 
is locally flat in those regions, or in other
words the density of states has van Hove singularities, such that
the single-particle transitions may turn into collective excitations
with a slightly different energy.  

For instance, for $2.8\omega_c<\omega<3.6\omega_c$, the peak at 
$\omega=2.95\omega_c$ corresponds to the energy interval $E_{3,0}-E_{0,0}=
2.77\omega_c$, plus a blue shift determined in part by the increased energy 
distance between the adjacent levels, $E_{3,X_0}-E_{0,X_0}$, with say 
$0<X_0<0.5/K$, and in part by the external electric field, similarly to the 
second term of Eq. (\ref{mdlq}).  The highest peaks, at $\omega=3.32\omega_c$ 
and $\omega=3.37\omega_c$, apart of this small splitting, can be related to 
the energy interval between the other pair of vHS, 
$E_{3,\pi/K}-E_{0,\pi/K}=3.27\omega_c$.  
The small splitting of the highest
mode cannot be explained only in terms of the energy spectrum.  The middle 
peak, at $\omega=3.18\omega_c$, may be put in correspondence with the
average energy interval, and thus considered as a combination of the two 
types of vHS modes.  Qualitatively, the same structure is obtained
for the Bernstein modes around the other multiples of $\omega_c$.

For $\nu=1.2$, the dashed line of Fig. 2, the states around the central 
vHS of the Brillouin zone become empty, and the corresponding subpeaks
vanish.  The bands are nearly flat, due to the screening effect, but the 
splitting of the higher subpeaks is still present.  Similar results
have been very recently obtained by Brataas et.\ al.\ \cite{Brataas:970311} 
for the singularities of the imaginary part of the dielectric function of 
modulated systems, by analytical calculations, without the inclusion of 
the screening in the groundstate, but also related to the nonparallel 
Landau bands.  

In order to support the above interpretation of the Bernstein subpeaks,
in Fig. 4 we show the particle density at equilibrium, $\rho(x)$, 
the induced density, $\rho_{ind}(x)$, and the extreme density profiles during 
the oscillations, $\rho(x)\pm\rho_{ind}(x)$, inside a unit cell,
$0\le x \le 2\pi$, for $\nu=2$ and $q=K$.  Obviously, the minima (maxima) 
of the equilibrium density correspond to the maxima (minima) of the 
modulation-potential energy. For graphical reasons we have amplified the 
induced density. The frequency of Fig. 4a, $\omega=2.70\omega_c$, is chosen
between two collective modes, where the absorption power is very
small.
The characteristic feature of such absorptionless oscillations is that
the evolution of the density preserves the reflection symmetry of the unit 
cell. A kind of a breathing mode, attributed to 
single-particle excitations in the presence of a dissipation mechanism. 

For the frequencies corresponding to collective peaks, Fig. 4b and c, we 
observe a global shift of the particle density inside the unit cell, i.\ e.\ 
we observe dipolar modes.  In Fig. 4b both the density minima, around $Kx=0$ 
and $Kx=2\pi$, and the maxima, around $Kx=\pi$, have horizontal oscillations
along the $x$ direction. The amplitude for the {\it minima} is 
larger than for the maxima, which is in agreement with the domination 
of this mode by the transitions around the {\it center} of the Brillouin zone
(associated to the maximum modulation-potential energy).  We also observe
a strong breathing component, due to the continuous spectrum of 
single-particle transitions of higher energies, which can be associated 
with the blue shift.  In Fig. 4c we have dipolar oscillations of the density
{\it maxima}, i.\ e.\ of frequency corresponding to the {\it lateral} vHS, 
with only a small breathing contribution, due to the excitation gap above.  
The neighboring mode, with $\omega=3.32\omega_c$, Fig. 2, is very similar 
to that of Fig. 4c, but having
an additional node in the induced density.  Also, the soft mode for $\omega= 
3.18\omega_c$ is similar to that of Fig. 4b, with one node less in the induced
density, but with a stronger breathing component. 

We will discuss now the situation when the modulation period is much larger
than the magnetic length.  In this regime the Landau bands are 
parallel, and the screening effects are stronger than before.  In Fig. 5 we
display the absorption for frequencies within the interval $0.5\omega_c<
\omega<3\omega_c$, again with a complicated structure of the Bernstein
modes.  As in the case with $a=50$ nm, this structure also reflects the 
many branches of the magnetoplasma oscillation spectrum,\cite{Wulf90:3113}
but now the internal peaks cannot be related to the energy spectrum as before,
because the excitation energies between the central and the lateral vHS 
coincide, e.\ g.\ like in Fig. 6.  
Here the  energy dispersion in the vicinity of 
the Fermi level is flat due to the electrostatic screening.\cite{Wulf88:4218a}
Since the states around the center of the Brillouin zone are less
populated than those at the edges, the active frequencies in the 
absorption spectrum correspond to dipolar oscillations of the density
maxima, as shown in Fig. 7.   The absorbant modes differ by the number of nodes
in the induced density.  In particular, for $\nu=1.6$, the height of the
absorption peaks is lower for the modes with fewer nodes, which is somewhat 
contrary to the expectation.  

A similar situation is shown in Fig. 8.  Here we have reduced the mean 
particle density to $\rho_0=0.5\times 10^{11}$ cm$^{-2}$, in order to explore 
the limit of isolated, parallel quantum wires.  The magnetic field is fixed 
to $B=4.67$ T. Even if the strips inbetween the wires are 
depleted to nearly zero particle density in the ground state, the wires 
still interact in the presence of the incident electric field. Consequently, 
the oscillations with a high number of nodes, capable of penetrating into the 
depleted strips, like in Fig. 8b, are energetically favorized with respect 
to those with less nodes, but which keep a rigid wire separation, Fig. 8c.  
Here we depict the density configurations for the middle and the right peaks 
in the group with three maxima of Fig. 8a.  For strictly isolated wires the 
induced density would be either symmetric or antisymmetric with respect to 
the wire center. This condition is the best obeyed by the mode at 
$\omega=1.26\omega_c$, which is thus specific to the wire response, 
the other modes being still under the influence of the electric coupling 
between the wires.

In Fig. 9a we display the evolution of the absorption spectra with increasing
density, from the limit of isolated wires, $\rho_0=0.1\times 10^{11}$ 
cm$^{-2}$, up to well coupled wires, $\rho_0=0.9\times 10^{11}$ cm$^{-2}$. 
The wave vector of the time dependent electric field is now small,
$q=0.1K$, in order to concentrate the attention only on the
effects of increasing the electron density on the absorption
in a periodic system. The effect of the gate, that we neglect here,
is mainly to repeat the same for each Bernstein mode.
The Fermi level is always in the Landau band with $n=0$.  For the lowest 
density the energy dispersion is large, but only few states are occupied at 
the bottom of the energy spectrum, while for the highest density the Landau 
band becomes very narrow, due to screening, such that the depleted strips 
vanish.  In the limit of isolated wires the induced density at the single 
absorption peak has only two nodes inside each unit cell, as can be seen in 
Fig. 9b.  This kind of motion has thus a strong breathing component, i.\ e.\ 
vertical oscillations of the total charge of the type shown in Fig. 4, and 
only small dipolar (horizontal) components which vanish with changing 
frequency.  Hence the internal motion of the electrons inside the wires 
has a strong contribution to the resonant modes, or in other words Kohn's 
theorem is not obeyed here, due to the nonparabolic, soft lateral 
confinement.  

Increasing the density, the wire coupling allows additional charge 
oscillations inside the unit cell, Fig. 9c, which may cumulate a stronger 
total dipolar component, yielding thus higher absorption peaks.  The lower 
the frequency, the higher the number of nodes within each wire, as also 
noticed for $q=K$, in Fig. 8.  The mode at $\omega=1.26\omega_c$, with few 
nodes per cell, corresponds to the only mode of the isolated wire system.
Such a mode is present in each trace of Fig. 9a, being the highest in 
frequency but the lowest in amplitude. Instead, the new modes, with 
additional internal oscillations, grow rapidly at lower frequencies, 
when the density, and thus the wire coupling, increases. 

For a given mode, the characteristic wavelength of the internal motions 
inside the unit cell is determined by the lateral confinement and by the 
wire coupling.  In our periodic system both of them result self-consistently 
from the external modulation potential, the magnetic field, and the electron 
density.  The evolution from a 
single-peak absorption around $\omega_c$, specific to isolated wires, to the 
multi-peak structure specific to the modulated system, occurs thus by the 
progressive appearance of new modes, with richer internal motion, and 
accumulating larger oscillator strength. For each density, the 
number of nodes in the absorption maxima of Fig. 9a gradually increases from 
right to left. When the density is increased further, beyond what is
shown in Fig. 9, such that more than one Landau band is occupied 
a stable {\it single peak emerges} with energy and induced density 
specific to the homogeneous system. 

\section{Summary}
In the present publication we have explored the properties
of the FIR-absorption of a 2DEG modulated in one direction,
varying the modulation and the electron density in order to
cover the whole range from isolated but electrically coupled
wires to a weakly modulated 2DEG. The electron-electron 
interaction has been treated self-consistently within the
time-dependent Hartree approximation. Experimentally the
transition from wires to an almost homogeneous 2DEG can be
achieved in heterostructures with a modulated metal gate,
which in turn also acts like a grating coupler for the 
incident FIR radiation. The modulation of the incident 
field should be determined self-consistently including
effects from the gate and the 3D structure of the 
sample.\cite{Lier94:7757}
To avoid this problem that is further complicated by the
presence of the magnetic field we considered primarily 
the effects of the two lowest Fourier components in the
incident field, ($q\approx 0$, $q=K$). The presence of the short wavelength
component and the magnetic field of intermediate strength
place the dispersion of the magnetoplasmon in the 
regime of Bernstein modes. 

Our calculations show that information about the 
nonparallel Landau bands in short period wires
can be extracted from the peak structure of the
FIR absorption. 

The effects of the vHS of the underlying energy
band structure on the dielectric function in weakly magnetically
and electrically modulated system have been studied in absence of the
Coulomb interaction between electrons.\cite{Stewart95:R17036,Stewart96:6019}
\acknowledgements
This research was supported by the Icelandic Natural Science Foundation, 
the University of Iceland Research Fund, and by the International Centre 
for Theoretical Physics, Trieste, Italy, within the Associateship Scheme.  
We thank Detlef Heitmann and Christoph Steinebach for helpful discussion.
One of us (A.M.) wishes to thank both University of Iceland and ICTP Trieste
for hospitality.


\begin{figure}
\epsfxsize 15cm 
\epsffile{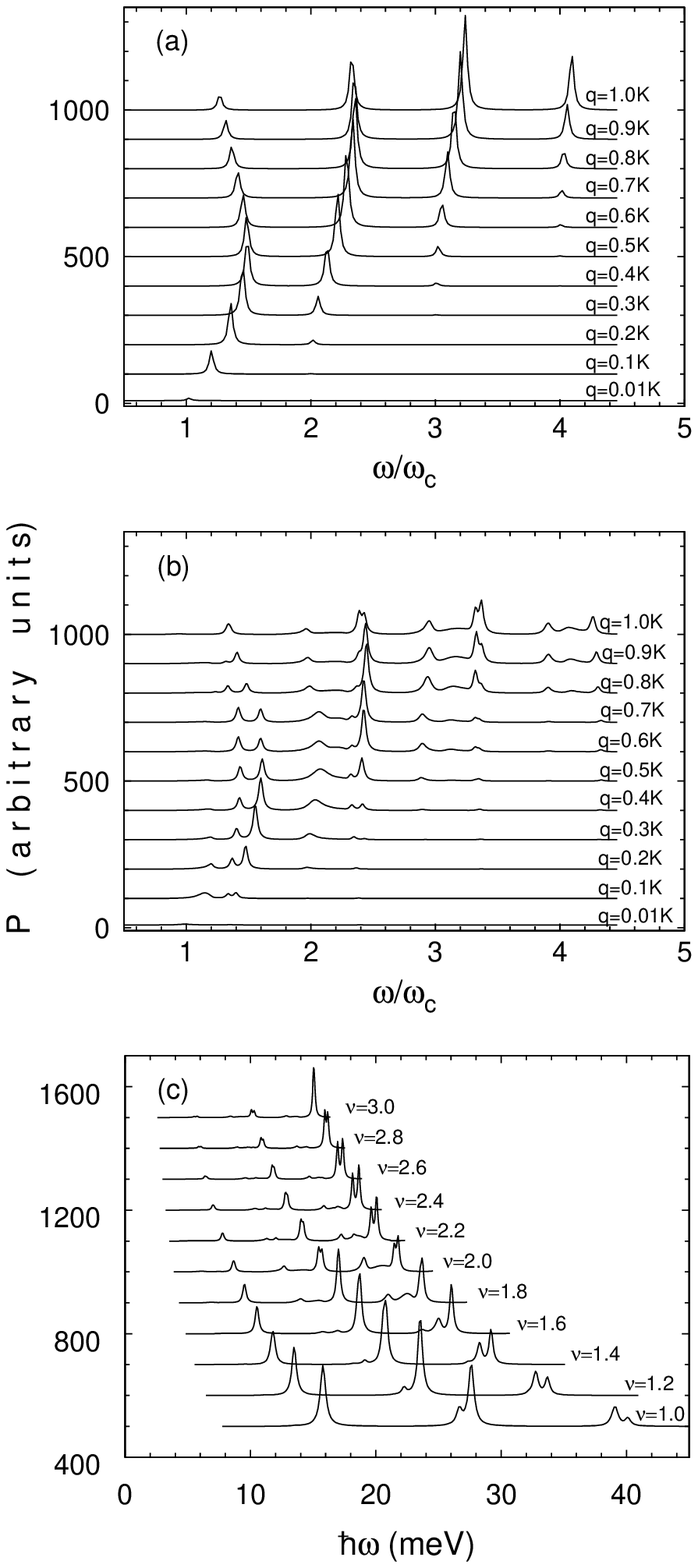}
\vspace{-4cm}
\caption{(a) Absorption spectra for a homogeneous system. 
$K=2\pi/a$, 
$a=50$ nm. The filling factor is 2 and the density 
$\rho_0=1.8\times 10^{11}$ cm$^{-2}$ (b) The same as (a), with a 
modulation
of period $a=50$ nm and amplitude $V=5$ meV. (c) The same modulation, 
with a
variable magnetic field such that the filling factor evolves between 1 
and 3. $q=K$.}
\end{figure}
\begin{figure}
\epsfxsize 12cm 
\epsffile{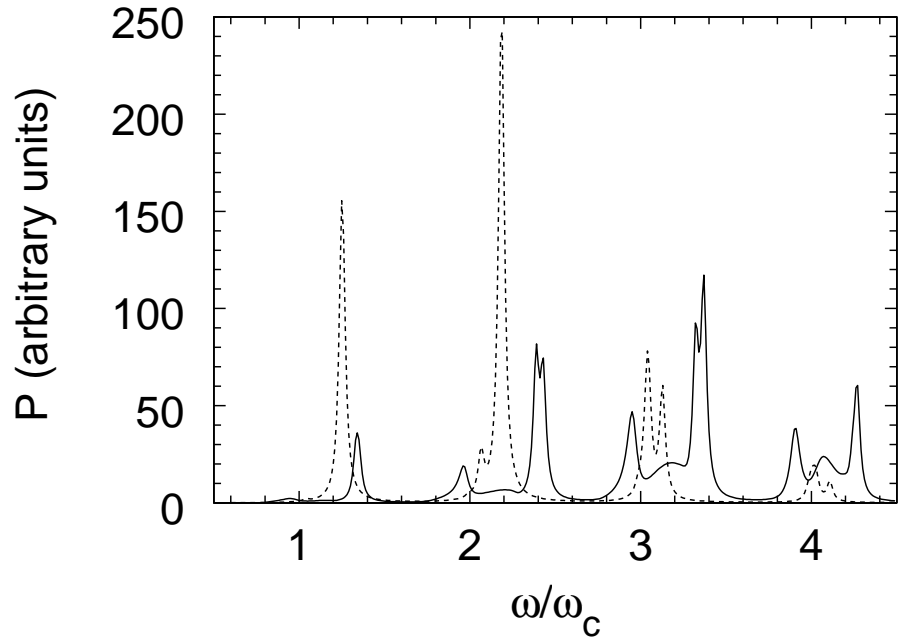}
\caption{The curves of Fig. 1c corresponding to $\nu=2$, full line, 
and $\nu=1.2$, dashed line, magnified.}
\end{figure}
\begin{figure}
\epsfxsize 12cm 
\epsffile{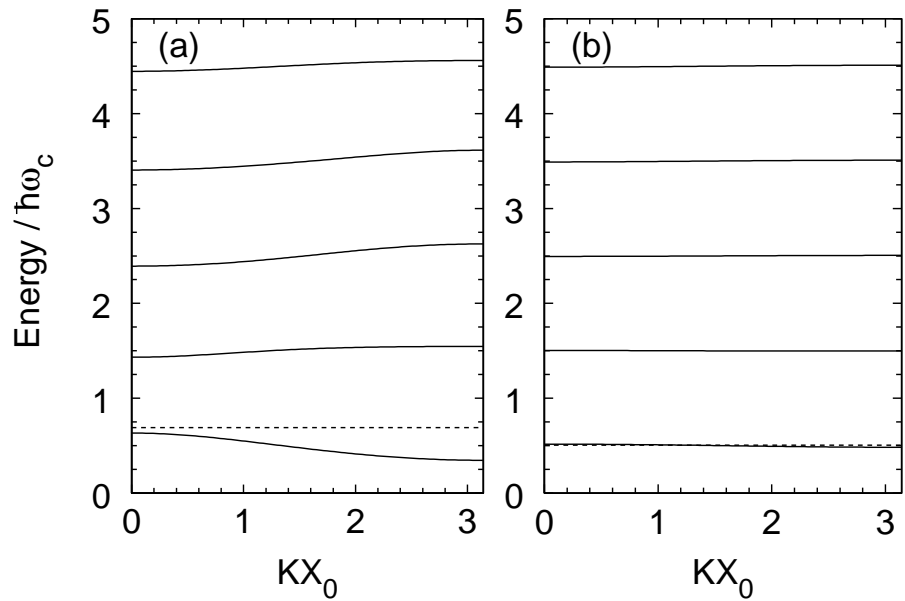}
\caption{Energy spectra for (a) $\nu=2$ and (b) $\nu=1.2$ in half of 
the
Brillouin zone.  The dashed horizontal lines show the Fermi level.}
\end{figure}
\begin{figure}
\epsfxsize 12cm 
\epsffile{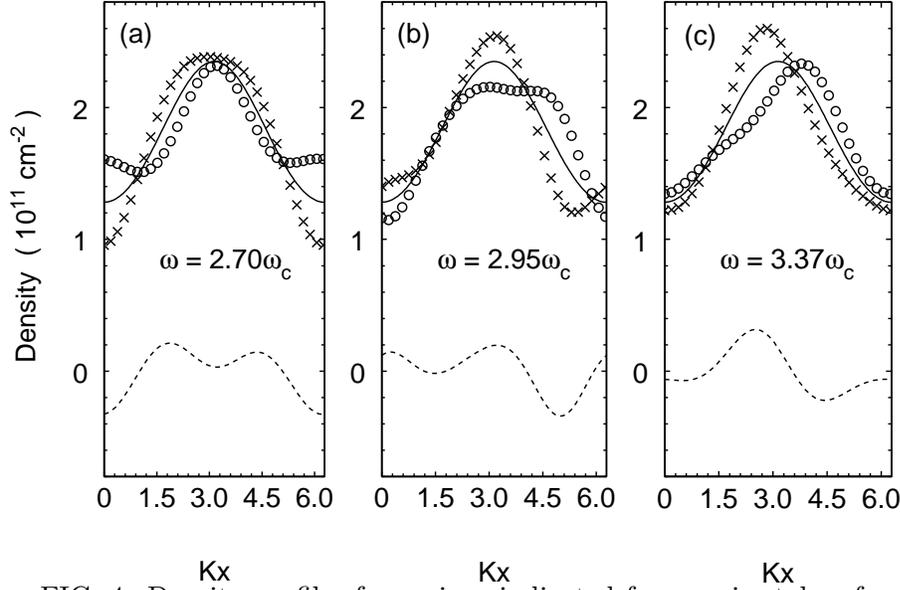}
\caption{Density profiles for various indicated frequencies taken from 
the
absorption spectrum of Fig. 2, $\nu=2$.  The full 
lines show the groundstate density, the dashed lines the maximal 
induced 
density, and the marked curves show the extreme density 
configurations.}
\end{figure}
\begin{figure}
\epsfxsize 12cm 
\epsffile{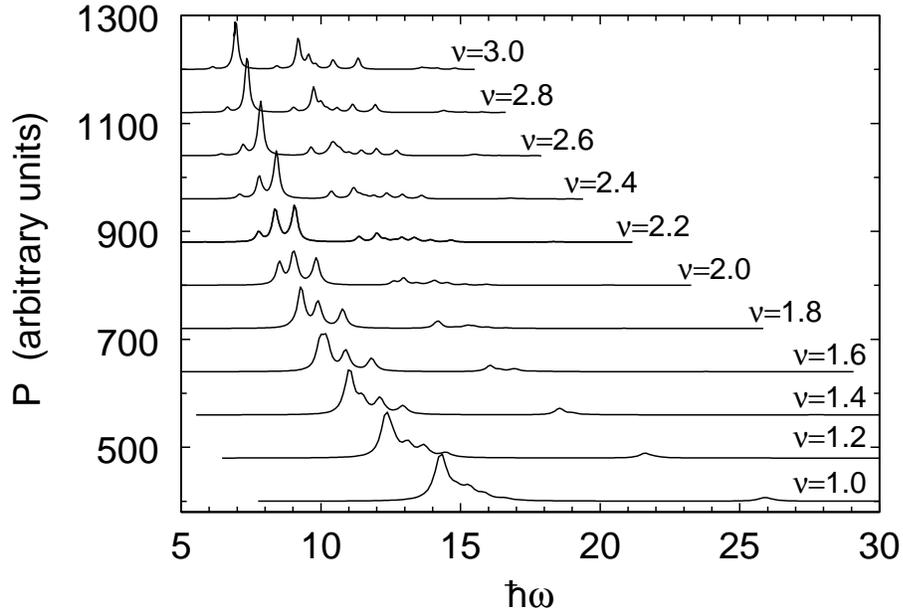}
\caption{Absorption spectra for a modulated system with $a=200$ nm 
and 
$V=20$ meV. The average density is $\rho_0=1.8\times 10^{11}$ 
cm$^{-2}$, $q=K$.}
\end{figure}
\begin{figure}
\epsfxsize 12cm 
\epsffile{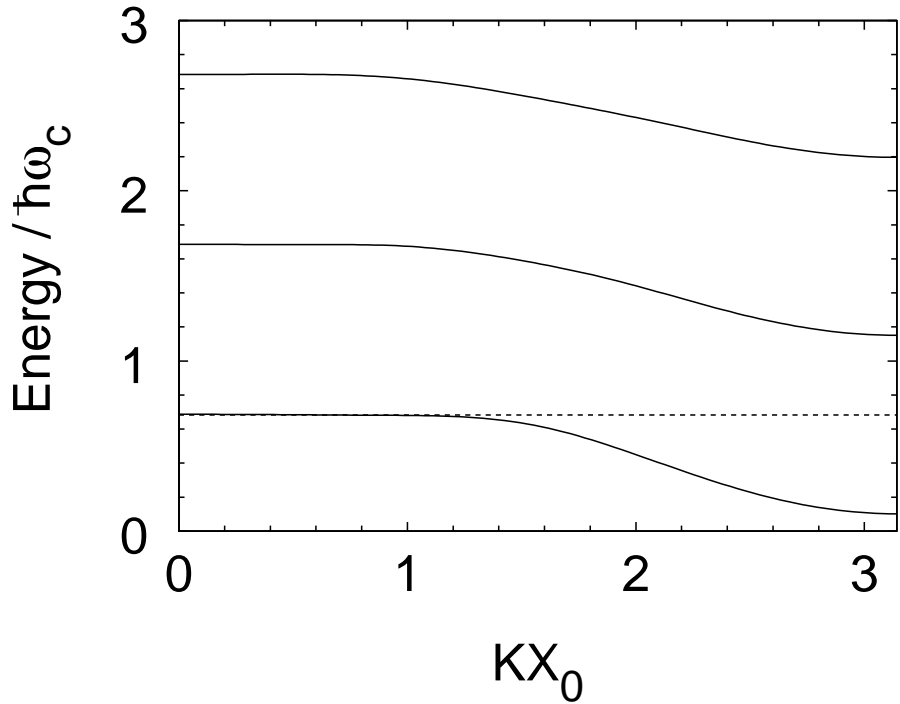}
\caption{Landau bands for the modulation of Fig. 5, with $\nu=1.6$.}
\end{figure}
\begin{figure}
\epsfxsize 12cm 
\epsffile{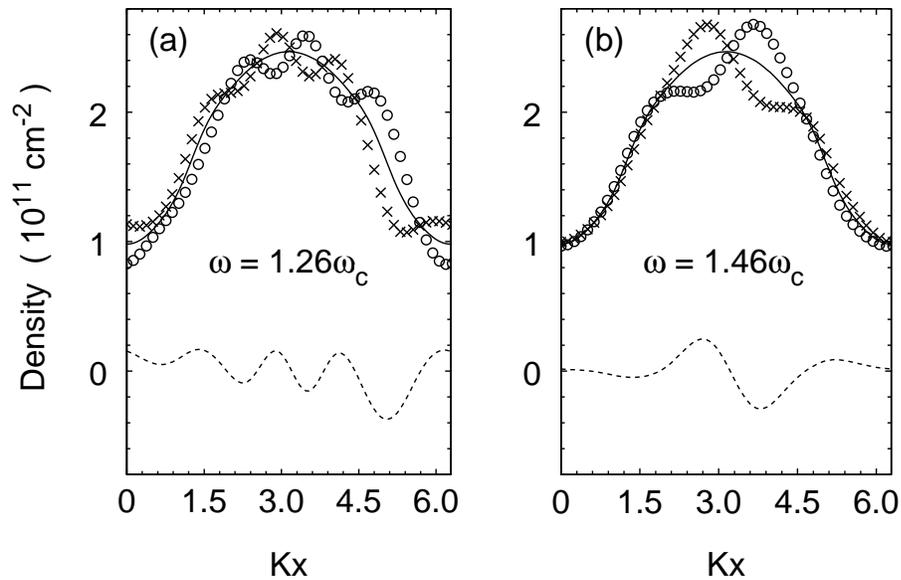}
\caption{Density profiles for oscillations in the absorption spectrum
of Fig. 5, for $\nu=1.6$. (a) $\hbar\omega=10.2$ meV, 
(b) $\hbar\omega=11.8$ meV.}
\end{figure}
\begin{figure}
\epsfxsize 15cm 
\epsffile{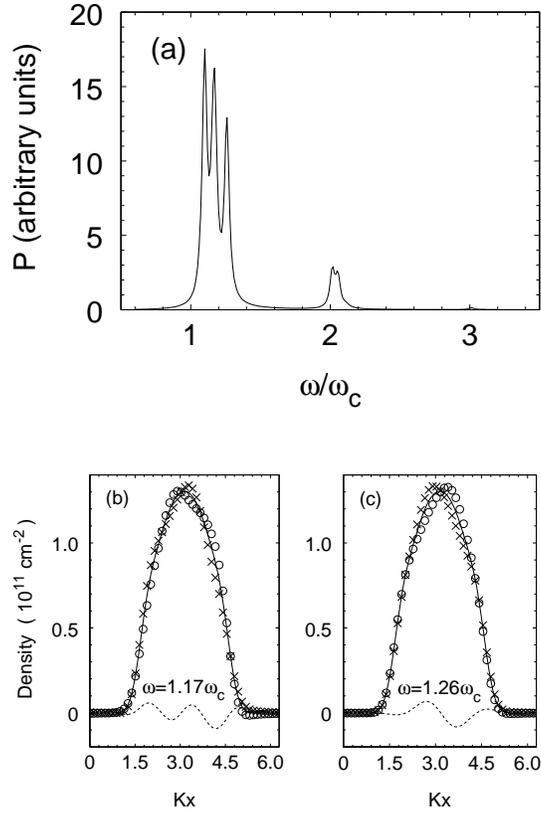}
\vspace{-6cm}
\caption{(a) Absorption peaks for the same modulation as for Fig. 5, 
but with
the average density $\rho_0=0.5\times 10^{11}$ cm$^{-2}$.  (b) and 
(c)
two density diagrams.}
\end{figure}
\begin{figure}
\epsfxsize 15cm 
\epsffile{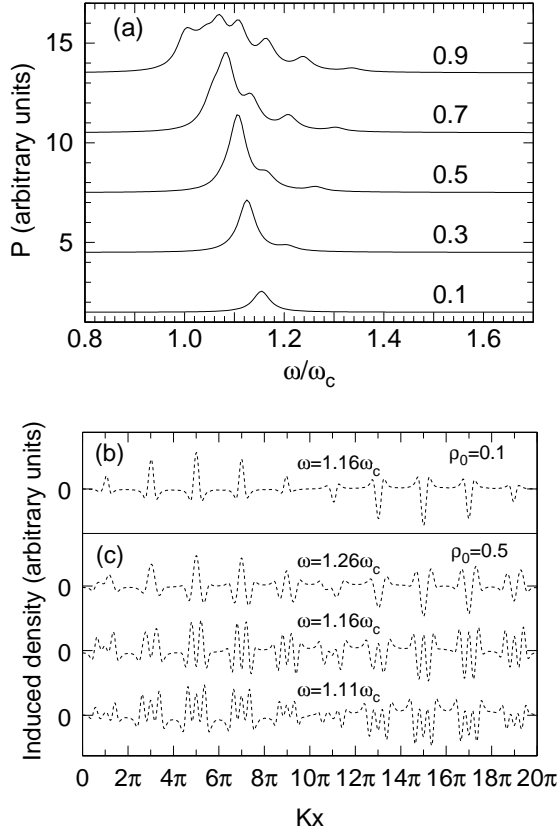}
\vspace{-6cm}
\caption{(a) Absorption spectra for various electron densities, from 
0.1 to 
0.9 units of 10$^{11}$ cm$^{-2}$, covering the transition from isolated 
wires 
to a modulated system.  The magnetic field is $B=4.67$ T and the 
wave vector
of the incident electric field is $q=0.1K$.  (b) and (c) The induced 
electron 
density, in a full period of the perturbed system, for 
$\rho_0=0.1\times 10^{11}$ cm$^{-2}$ and $\rho_0=0.5\times 
10^{11}$ cm$^{-2}$, 
for the frequencies corresponding to the absorption maxima.}
%
\end{figure}
\end{document}